\documentclass[final]{aipproc}

\layoutstyle{6x9}

\begin{document}

\title{Nuclear Anapole Moments and the 
       Parity-nonconserving Nuclear Interaction}

\author{Cheng-Pang Liu} 
{address={TRIUMF Research Facility,  
          4004 Wesbrook Mall, Vancouver, BC, Canada V6T 2A3}}

\begin{abstract}

The anapole moment is a parity-odd and time-reversal-even
electromagnetic moment. 
Although it was conjectured shortly after the discovery of parity 
nonconservation, its existence has not been confirmed until 
recently in heavy nuclear systems, which are known to be the
suitable laboratories because of the many-body enhancement. 
By carefully identifying the nuclear-spin-dependent atomic parity
nonconserving effect, the first clear evidence was found in 
cesium. 
In this talk, I will discuss how nuclear anapole moments are 
used to constrain the parity-nonconserving nuclear force, a 
still less well-known channel among weak interactions.

\end{abstract}

\maketitle

\subsection{Introduction}

Tests of the unified electroweak theory have long been an 
important subject in physics. 
Compared with successes gained in the leptonic, semileptonic, 
and flavor-changing hadronic sectors, it is fair to say that 
the flavor-conserving hadronic weak interaction is not 
well-constrained.
Experimentally, this sector could only be studied in nuclear 
systems, therefore, the major challenge comes from the 
sensitivity required to separate the parity-nonconserving (PNC) 
observables from much larger strong and electromagnetic (EM) 
backgrounds. 
Despite a number of difficulties, these studies are of 
fundamental importance because the hadronic neutral weak 
interaction only appears in the flavor-conserving sector. 
One also hope that these studies can reveal more information 
about the hadronic dynamics which can not be probed by 
parity-conserving (PC) observables. 

Several precise and interpretable nuclear PNC measurements 
have already been made and put constraints on the PNC 
nucleon-nucleon (NN) interaction. 
These results along with new developments using polarized proton 
or neutron beam will be reviewed in the plenary talk by W. D. 
Ramsay.
The focus of this short presentation is the nuclear anapole 
moment, which provides another window to examine the PNC NN 
interaction and have to be measured in atomic PNC experiments.

It was first noted independently by Vaks and Zel'dovich 
\cite{Ze58} that the PNC mechanism allows a parity-odd EM 
coupling to an elementary particle (actually to a composite 
system as well) by inducing an exotic electromagnetic moment, 
called the ``anapole moment'' (AM).  
Later on, Flambaum, Khriplovich, and Sushkov \cite{FKS84} pointed 
out that the AM of a nucleus grows roughly as the nucleon number 
to the power of two-thirds, and this suggested it might be 
possible to measure this nuclear AM in heavy atoms.

These theoretical conjectures finally realized when Colorado 
group \cite{Colorado97} announced the first clear evidence of 
nuclear AM using the polarized atomic cesium beam. 
By carefully identifying the hyperfine dependence of atomic PNC 
effects, a 7$\sigma$ determination was reported. 
And the error bar is so small that a very good constraint 
on the PNC NN interaction could be deduced (if one does the 
calculation right). 
In the context of this symposium, it is more than adequate to 
recognize this discovery by atomic physicists which contributes 
to the nuclear spin physics at the lowest energy end.

\subsection{Nuclear AMs and the PNC NN Interaction}

According to the multipole expansion, the electromagnetic moments 
are classified as charge $C_J$, transverse electric $E_J$, and 
transverse magnetic $M_J$, where $J$ denote the angular momentum. 
Normally, parity (P) and time-reversal (T) invariance only allow 
charge moments of even order ($C_0$, total charge; $C_2$, charge 
quadrupole; ...etc.) and transverse magnetic moments of odd order 
($M_1$, magnetic dipole; $M_3$, magnetic octupole; ...etc.). 
The vector moment $C_1$, which is P- and T-odd, is the charge 
dipole moment. 
The vector moment $E_1$, which is P-odd but T-even, is exactly 
the ``anapole moment''. 
Often in the literature, the anapole operator $\vec{a}$, 
generated by the current density operator $\vec{j}(\vec{r})$, 
is defined as

\[
\vec{a} = - \pi \int d^3 r \, r^2 \vec{j}(\vec{r}) 
        \equiv \frac{G_F}{\sqrt{2}} \, \kappa_{am} \, \vec{I}.
\footnote{The current conservation plays a role in defining 
the form of $E_1$ operator, and the definition for $\kappa$ is 
different from what Khriplovich \textit{et al.} adopted. 
For more details, see Ref. \cite{HLM02}.}
\]
It is clear that this operator gives vanishing expectation values 
unless the wave function is not a parity eigenstate or the 
current is axial-vector---both are linked to weak interactions at 
the fundamental level. 
In the last part of the equation above, a dimensionless quantity 
$\kappa_{am}$, which characterizes the strength of a nuclear AM, 
is defined through the Fermi constant (the typical scale of weak 
interaction) and the nuclear spin vector $\vec{I}$ (the only 
intrinsic vector of an elementary or composite particle).

An illustrative picture of the AM is the toroidal current winding. 
Because the $r^2$ weighting factor, the currents on the outer part 
of the torus give larger moments than the inner part, and this 
leads to a net AM. 
Suppose in a system where parity is a good symmetry, the left- 
and right-handed current windings should be equally probable, 
therefore no net AM occurs. 
However, any non-equal mixture of these two by some PNC mechanism 
will results in a chiral current and thus an AM. 
Also noteworthy is that the magnetic field generated by the
toroidal current winding is confined, therefore, unless a 
particle is inside the torus, there in no interaction. 
This contact character of interactions with AMs is the same as 
the low energy neutral weak interaction, a result anticipated by 
the unified electroweak theory.

Although one believes the nuclear AMs have their origin in the 
couplings of quarks and weak bosons, $W^{\pm}$ and $Z^0$, 
a hadronic theory from the first principle is still unavailable. 
Instead, various models are designed to describe the 
nonperturbative dynamics of hadrons. 
For the PNC NN interaction, the widely-used framework is a one 
meson exchange model including $\pi$, $\rho$, and $\omega$, 
with one of the meson couplings is PC and the other PNC.
The six PNC meson coupling constants in this model,
$h_{\pi}$, $h_{\rho}^0$, $h_{\rho}^1$, $h_{\rho}^2$, 
$h_{\omega}^0$, and $h_{\omega}^1$, as defined by Desplanques, 
Donoghue, and Holstein (DDH) \cite{DDH},
undermine the physics of how the 
fundamental couplings of quark and weak bosons are modified by 
the strong interaction.
\footnote{$h_{\pi}$ was named ad $f_{\pi}$ originally by DDH, 
however, it is changed in order not confuse with pion decay 
constant sometimes.} 
The theoretical benchmark is given by the so-called DDH ``best 
values'' along with some reasonable ranges. 
It is the hope that experiments could constrain these couplings 
well and justify the hadronic theories.

Given this PNC NN potential, nuclear AMs arise in three ways:
i) one-body contribution, where weak radiative corrections are 
induced in the form of single nucleon loop or pole diagrams, 
often called as the nucleonic AM, 
ii) two-body contribution, where mesons induce extra EM 
currents by coupling photons to nucleon-antinucleon pairs and 
mesons in-flight,  
iii) polarization mixing, where a parity eigenstate state is 
mixed by opposite-parity states, thus the normally forbidden 
EM couplings are allowed.
While the one-body contribution is incoherent, many-body effects 
would possibly enhance the two-body and polarization mixing 
terms.

\subsection{Experimental Results and Deduced Constraints}

With the increasing accuracy, atomic PNC experiments have been an 
important part of the low-energy precision tests of the standard 
model. 
The dominant PNC effect comes from the tree-level $Z^0$ exchange 
between atomic electrons and the nucleus with an axial-vector 
coupling at electrons and a vector coupling at the nucleus, 
A(e)-V(N). 
This is a nuclear-spin-independent (NSI) effect in which every 
nucleon contributes coherently. 
On the other hand, The V(e)-A(N) exchange gives a 
nuclear-spin-dependent (NSD) effect, but is much suppressed 
because nucleons contribute incoherently and electrons have a 
weaker vector coupling to $Z^0$.

Although the interaction of electrons with the nuclear AM comes 
at a higher order, i.e., $G_F\alpha$, it actually dominates the 
NSD effect in heavy nuclei because the electron coupling is not 
suppressed by (1 - 4 Sin$^2\theta_W$) and the nuclear many-body 
enhancement grows as $A^{2/3}$. 
Therefore, the extraction of nuclear AM involves: 
i) an atomic many-body calculation relating the experiment result 
to the PNC electron-nucleus interaction, 
ii) the identification of NSD PNC effect by comparing results from 
different hyperfine levels, and 
iii) the subtraction of contributions due to $Z^0$ exchange and 
hyperfine interaction.

So far, nuclear AMs in cesium and thallium have been reported. 
The cesium experiment by Colorado group showed a clear evidence, 
however, the thallium experiments by Seattle \cite{Seattle95} 
and Oxford \cite{Oxford95} groups had large error bars so the 
results are consistent with zero. 
The extracted AMs in terms of $\kappa_{am}$ are: 
$\kappa_{am}$(Cs) = 0.090 $\pm$ 0.016 \cite{FlMu97} and 
$\kappa_{am}$(Tl) = 0.376 $\pm$ 0.400 (Seattle's only).
\footnote{The Oxford result is not quoted here, see Ref. 
\cite{HLM02} for discussion.}

In order to constrain the PNC meson couplings using these
results, one has to perform a model calculation of the nuclear 
AM and then express $\kappa_{am}$ in terms of these couplings. 
Because both Cs and Tl are heavy nuclei, the nuclear structure 
is the most important issue. 
There have been quite a few calculations with various treatments, 
a brief summary and survey could be found in Ref. \cite{DmKh02}. 
Roughly speaking, the calculations based on the single particle 
approximation, which treat the Cs as a $1g_{7/2}$ proton 
plus the closed core and Tl as a $3s_{1/2}$ proton hole 
plus the closed core, tend to predict larger AMs than 
calculations which consider many-body effects.

The constraints on the PNC meson couplings is presented in 
Fig. 1. 
The Cs and Tl bands are plotted based on the shell model results 
of Ref. \cite{HLM01,HLM02}, a full two-body calculation in which 
all the exchange currents are included and the polarization 
mixing is handled by the closure approximation with the aid of 
nuclear systematics found in light nuclei. 

\begin{figure}
{\centering\resizebox*{0.5\columnwidth}{!}
{\includegraphics{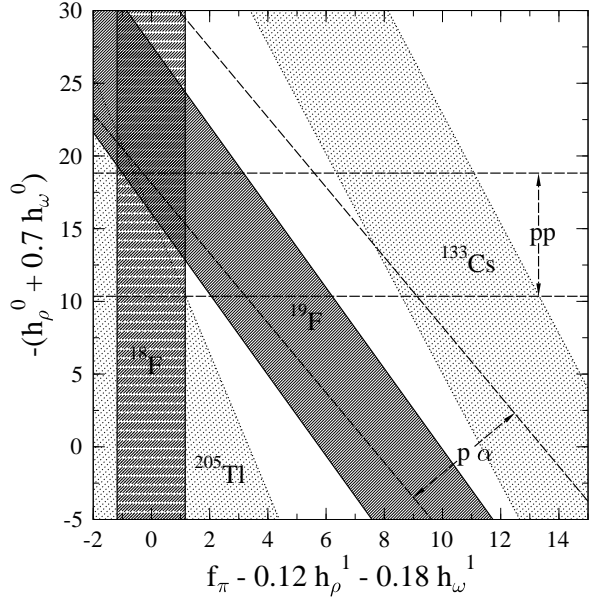}}\par}
\caption{Constraints on the PNC meson couplings ($\times 10^7$). 
The error bands are one standard deviation.}
\end{figure}

Apparently, the anapole constraints are not in agreement with the 
existing nuclear PNC results, and also with each other (only a 
small part of the Tl band is shown here, and the central line of 
this band has a negative $x$-intercept). 
The result for the AM of Tl is rather confusing because the 
experiment gave a positive value, but all the calculations 
predict negative. 
Therefore, it is very possible that the tension between Cs and 
Tl bands is due to this sign problem.
One can also observe that the Cs result tests a similar 
combination of PNC couplings as $p \alpha$ and $^{19}$F, but 
favors lager values. 
By combining Cs and $pp$ bands, the allowed region does fall into 
the DDH reasonable ranges, with $h_{\pi}\sim$ 9. 
However, the stringent limit set by the $^{18}$F result, 
$|h_{\pi}|\le$ 1, which has been performed by five groups, 
definitely rules out this possibility.

The big discrepancy between the anapole constraints and existing 
nuclear PNC results is certainly a puzzle to be sorted out. 
The first criticism of the theory would be on our still limited 
knowledge of the structure of heavy nuclei.
By the way, the atomic many-body theory, which is the key to the 
interpretation of experiments, should also be examined. 

With only one certain result in Cs, obviously we need more 
experimental inputs to clarify the current situation. 
There are several new measurements being in progress or 
proposal. 
For example, a new Cs measurement will double-check the existing 
result, 
an improved Tl experiment hopefully can solve the sign problem, 
results of odd-neutron nuclei like Dy, Yb, and Ba would produce 
constraints roughly perpendicular to what odd-proton nuclei do, 
and the study of a chain of Fr isotopes should reduce some of the 
theoretical uncertainties.

However, it ought to be emphasized that, if any of these 
results, when available, is going be to used for constraining PNC 
meson couplings reliably, a good nuclear structure calculation 
is still the top necessity.

\subsection{Summary}

The nuclear anapole moment, a manifestation of nuclear
parity-nonconservation which has been conjectured for a long 
time, is clearly discovered in the atomic PNC experiment of 
cesium.
The precision of this result makes it sensible to constrain the 
PNC meson couplings. 
However, a big discrepancy is found by comparing this new 
constraint with existing nuclear PNC results, most possibly due 
to the nuclear structure uncertainties. 
In order to constrain the hadronic theory reliably, this puzzle 
should be further addressed.

\begin{theacknowledgments}

The author would like to thank Profs. W. van Oers and M. J. 
Ramsey-Musolf for encouraging this presentation at 
SPIN 2002 symposium.

\end{theacknowledgments}

\end{document}